\documentclass[aps,twocolumn,amsmath,amssymb,showpacs]{revtex4-1}
\usepackage[ansinew]{inputenc}
\usepackage{graphicx}
\usepackage{xcolor}
\usepackage{siunitx}
\usepackage{hyperref}

\DeclareSIUnit\ions{ions}

\begin{document}

\title{Erbium emitters in commercially fabricated nanophotonic silicon waveguides}

\author{Stephan Rinner}
\email{These authors contributed equally to this work.}
\author{Florian Burger}
\email{These authors contributed equally to this work.}
\author{Andreas Gritsch}
\email{These authors contributed equally to this work.}
\author{Jonas Schmitt}
\author{Andreas Reiserer} 
\email{andreas.reiserer@tum.de}

\affiliation{Max-Planck-Institut f\"ur Quantenoptik, Quantum Networks Group, Hans-Kopfermann-Strasse 1, D-85748 Garching, Germany}
\affiliation{Technical University of Munich, TUM School of Natural Sciences and Munich Center for Quantum Science and Technology (MCQST), James-Franck-Stra{\ss}e 1, D-85748 Garching, Germany }

\begin{abstract}
Quantum memories integrated into nanophotonic silicon devices are a promising platform for large quantum networks and scalable photonic quantum computers. In this context, erbium dopants are particularly attractive, as they combine optical transitions in the telecommunications frequency band with the potential for second-long coherence time. Here we show that these emitters can be reliably integrated into commercially fabricated low-loss waveguides. We investigate several integration procedures and obtain ensembles of many emitters with an inhomogeneous broadening of $\lesssim\SI{2}{\giga\hertz}$ and a homogeneous linewidth of $\lesssim\SI{30}{\kilo\hertz}$. We further observe the splitting of the electronic spin states in a magnetic field up to \SI{9}{\tesla} that freezes paramagnetic impurities. Our findings are an important step towards long-lived quantum memories that can be fabricated on a wafer-scale using CMOS technology.
\end{abstract}

\maketitle

\section{Introduction}

A long-lived quantum memory for light is an essential ingredient for the development of large-scale quantum networks \cite{wehner_quantum_2018} and all-optical quantum computers \cite{nunn_enhancing_2013, makino_synchronization_2016}. In this context, ensembles of rare-earth dopants are particularly promising, as they allow for the creation of efficient memories \cite{tittel_photon-echo_2010, hedges_efficient_2010} that can have exceptional coherence when storing excitations in the spin of such dopants \cite{zhong_optically_2015, rancic_coherence_2018}. Most landmark experiments on this platform, such as the storage of entanglement in separate crystals \cite{clausen_quantum_2011, saglamyurek_broadband_2011, lago-rivera_telecom-heralded_2021}, used bulk crystals. For scaling up, however, it is desirable to use nanophotonic devices that have a small footprint and can be fabricated in large numbers, which is required for the parallel operation of multiple devices. Pioneering steps in this direction used rare-earth emitters in photonic waveguides \cite{marzban_observation_2015, falamarzi_persistent_2020, rakonjac_storage_2022, liu_-demand_2022} and nanophotonic cavities \cite{zhong_nanophotonic_2017, dibos_atomic_2018, craiciu_multifunctional_2021, ourari_indistinguishable_2023} made from the same host crystals as used in the bulk experiments, such as yttrium orthovanadate~\cite{zhong_nanophotonic_2017}, yttrium orthosilicate ~\cite{dibos_atomic_2018, craiciu_multifunctional_2021, liu_-demand_2022} or calcium tungstate~\cite{ourari_indistinguishable_2023}. Unfortunately, these and related materials are not compatible with wafer-scale manufacturing using established processes of the semiconductor industry. A promising alternative is therefore to integrate rare-earth dopants directly into silicon, where narrow inhomogeneous~\cite{weiss_erbium_2021} and homogeneous~\cite{gritsch_narrow_2022} linewidths have been observed, and single dopants have been resolved in a nanophotonic resonator~\cite{gritsch_purcell_2023}. These experiments used the element erbium which has a coherent optical transition at a telecommunication wavelength~\cite{bottger_optical_2006, merkel_coherent_2020}, around \SI{1536}{\nano\meter}, where absorption in silicon is negligible and optical fibers exhibit minimal loss. The nanophotonic devices in previous publications \cite{weiss_erbium_2021, gritsch_narrow_2022, gritsch_purcell_2023} were fabricated using electron-beam lithography and reactive-ion etching after the entire device layer of small silicon-on-insulator chips had been implanted with erbium. In this work, we demonstrate that similar results can be obtained with silicon waveguides that were fabricated commercially on a multi-project-wafer (MPW) and implanted afterwards. We investigate which is the best erbium integration procedure and characterize the optical and spin properties of the emitters. This constitutes an important step towards silicon-based quantum memories that are compatible with mass fabrication.

\begin{figure}[hb!]
\includegraphics[width=1\columnwidth]{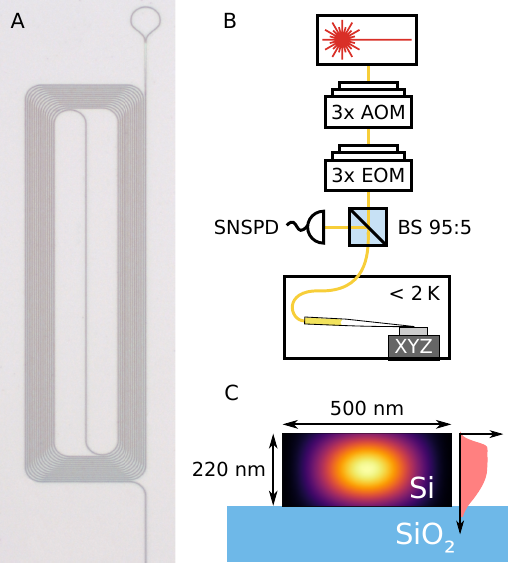}
\caption{\label{fig:Setup} A) Optical microscope image of one of the investigated structures. The waveguide is coiled up to maintain a small footprint. One end is terminated by a loop-shaped retroreflector (top), the other end (not shown) is used for in- and outcoupling of light. B) Measurement setup. A laser is used to excite the dopants. It is pulsed and frequency-shifted using acousto-optical modulators (AOM). Furthermore, sidebands can be generated with three concatenated electro-optical modulators (EOM). The light is coupled into the waveguides on the chip using a tapered fiber on a nanopositioning stage in a cryostat. Before, it passes a fiber-optical 95:5 beam splitter, which attenuates the input and guides the backward-propagating light predominantly to a superconducting nanowire single photon detector (SNSPD) mounted in another cryostat. C) Waveguide cross section. The waveguide consists of crystalline silicon (black) on top of a silicon dioxide layer (blue). It supports a single guided TE-like mode (the color gradient shows the flux density). Erbium was implanted into the waveguides at three different combinations of implantation dose and energy, resulting in an approximately homogeneous vertical erbium distribution (right side, light red).}
\end{figure}

\section{Samples and setup}

The silicon photonic chips were fabricated in an MPW run at \textit{Advanced Micro Foundry} (AMF), and subsequently implanted with $^{170}\text{Er}$ under an angle of $\SI{7}{\degree}$ and at different temperatures by \textit{Ion Beam Services} (IBS). Each sample was implanted sequentially with three parameter sets to achieve an approximately homogeneous profile of the erbium distribution in the waveguide: with a dose of \SI{5e11}{\ions\per\centi\meter\squared} at \SI{70}{\kilo\electronvolt}, \SI{9e11}{\ions\per\centi\meter\squared} at \SI{160}{\kilo\electronvolt}, and \SI{22e11}{\ions\per\centi\meter\squared} at \SI{350}{\kilo\electronvolt}. On one end, the waveguides are tapered, which allows for broadband coupling with a tapered single-mode optical fiber \cite{weiss_erbium_2021} that may be further improved in the future by dielectric coatings \cite{khan_low-loss_2020}. The other end is a loop reflector, which is formed by connecting the output arms of a balanced beam splitter with a waveguide bend, as shown in Fig. \ref{fig:Setup}A. In between, the waveguide is formed as a spiral to achieve a long length at a compact footprint. The samples are mounted in a closed-cycle cryostat with a variable temperature down to $<\SI{2}{\kelvin}$. We then perform resonant spectroscopy using the setup in Fig. \ref{fig:Setup}B. The waveguide cross section is \qtyproduct[product-units=single]{500 x 220}{\nano\meter}, as shown in panel C together with the simulated implantation profile.

\section{Erbium integration}

The experiments investigate large ensembles of erbium dopants in sites A and B, which emit at \SI{1537.76}{\nano\meter} and \SI{1536.06}{\nano\meter}, respectively, and have previously shown promising optical properties: narrow inhomogeneous and homogeneous broadening, and short lifetimes that are predominantly radiative \cite{gritsch_narrow_2022}. If erbium dopants are integrated into these or one of the many other possible sites \cite{berkman_observing_2023} that typically have less favorable properties depends on the purity of the silicon that forms the waveguide, and on the implantation and annealing conditions \cite{gritsch_narrow_2022}. Here, one expects a trade-off: At higher temperatures, the crystal damage caused by the implantation will be reduced, leading to a lower waveguide loss. In addition, a narrower inhomogeneous broadening of the optical transition frequency of the preferred erbium sites may be observed in case the strain inhomogeneity is reduced by the annealing. However, the mobility of both erbium and other impurities in the crystal also increases with temperature. They may thus form clusters, yielding many possible other site configurations. 

\begin{figure*}[ht!]
\includegraphics[width=2\columnwidth]{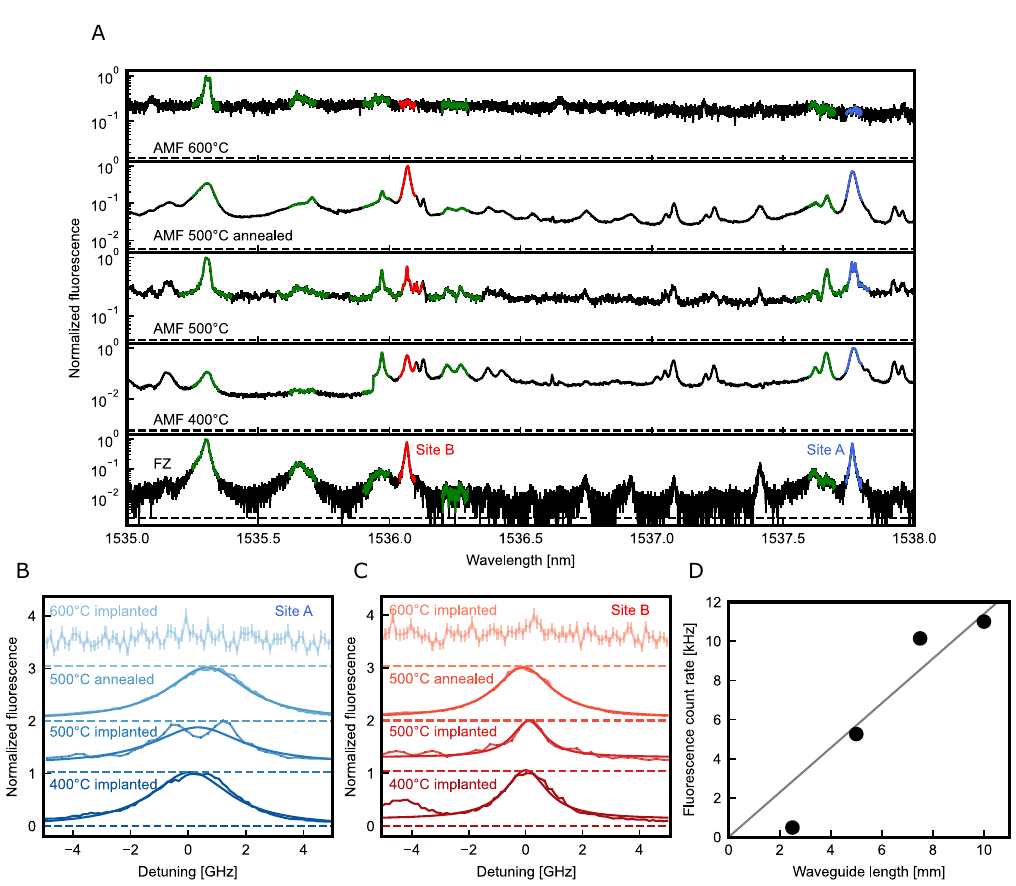}
\caption{\label{fig:BroadSpectrum}
A) Fluorescence scan. The frequency of the excitation laser is swept, and the fluorescence is recorded in a \SI{0.5}{\milli\second} time interval after the excitation pulse. The lowest curve shows the spectrum obtained on home-made float-zone silicon waveguides for comparison. Several different implantation conditions are studied on commercially fabricated samples (bottom to top): implanted at $\SI{400}{\celsius}$, $\SI{500}{\celsius}$, at room temperature with a $\SI{500}{\celsius}$ post-implantation anneal, and implanted at $\SI{600}{\celsius}$. Efficient integration into the desired erbium sites A and B (red and blue area) is achieved below $\SI{600}{\celsius}$. B) and C) Inhomogeneous broadening of the emission from the two preferred erbium sites. The linewidth is measured by scanning the excitation laser frequency with a high resolution. D) Obtained signal as a function of waveguide length, measured on the same sample. The clear increase up to $\SI{10}{\milli\meter}$ length suggests small loss in the implanted structures.
}
\end{figure*}

To investigate the integration into sites A and B in the commercially-fabricated chips studied in this work, we use the same techniques as in our earlier work \cite{gritsch_narrow_2022}. Specifically, we measure pulsed fluorescence, i.e. the signal emitted after exciting the dopant ensemble with resonant laser pulses of \qtyrange{0.1}{0.25}{\milli\second} duration. The pulses have a rectangular temporal profile, generated with acousto-optical modulators with a rise time on the order of $\SI{10}{\nano\second}$. This is more than four orders of magnitude shorter than the emitter lifetime, which allows separating the emission from the excitation. We first perform a broad scan of the excitation laser frequency, which is slightly modulated to avoid effects of persistent hole burning. The result of this measurement for several implantation and annealing conditions is shown in Fig. \ref{fig:BroadSpectrum}A. For comparison, we also include a spectrum obtained in our earlier measurements~\cite{gritsch_narrow_2022} on high-purity float-zone (FZ) silicon waveguides, implanted at room temperature and then annealed at \SI{500}{\celsius} for approximately one minute, on which waveguides were fabricated by our group. These chips will be referred to as `home-made' in this work.
 
We find that at \SI{400}{\celsius} erbium is efficiently integrated into the preferred sites A and B (marked blue and red, respectively), but also into pair- or precipitate sites (green) observed previously. In addition, we observe several other peaks and a broadband background. A very similar spectrum is obtained in the samples that were implanted at $\SI{500}{\celsius}$. In contrast, below $\SI{400}{\celsius}$ the optical loss is significantly higher. This prevents coupling into waveguides implanted at room temperature. Still, if these chips are subjected to a rapid thermal annealing at \SI{500}{\celsius} for \SI{1}{\minute}, we again obtain low losses and a high integration yield. Finally, at an annealing temperature of \SI{600}{\celsius} the fluorescence signal of sites A and B disappears completely, and the only remaining line originates from erbium precipitate sites (green)~\cite{gritsch_narrow_2022}, which suggests that at this implantation temperature the mobility of erbium and other impurities is too high such that clusters of several dopants and/or impurities are formed.

After characterizing the spectrum of the erbium emission over a broad range, we determine the inhomogeneous broadening of the emission from dopants in the two preferred sites. A corresponding high-resolution scan of the excitation frequency is shown in Figs. \ref{fig:BroadSpectrum}B and C. Some difference of the linewidths is observed; however there is no strong dependence on the implantation conditions. We find that a Lorentzian distribution fits our data the best. On the sample annealed at \SI{500}{\celsius}, we extract full-width-at-half-maximum (FWHM) linewidths of \SI{3.49 \pm 0.18}{\giga\hertz} for site A and \SI{1.87 \pm 0.09}{\giga\hertz} for site B. This is significantly larger than the values of $\sim\SI{0.5}{\giga\hertz}$ obtained in our previous experiments that were performed on samples grown by the Czochralski technique (obtained from Silicon Valley Microelectronics) and implanted at comparable parameters \cite{gritsch_narrow_2022}. This suggests that the silicon device layer in the commercially fabricated chips has a larger strain inhomogeneity. This may originate directly from the used wafer (of proprietary origin) or be caused by impurities and defects added in the fabrication, implantation and/or annealing processes. To clarify, further investigations will be required. While the increased inhomogeneous broadening may enhance the bandwidth of an ensemble-based memory \cite{tittel_photon-echo_2010}, it also reduces the optical depth, so there will be a trade-off between efficiency and bandwidth in future memory experiments.

In addition to the resonant emission, we also observe an offset, i.e. a broad fluorescence background, that stems from dopants that can be excited off-resonantly, similar to our earlier work \cite{gritsch_narrow_2022}. This background is more pronounced in the commercial samples studied here. It can be reduced by inserting a narrow-band filter that is set to transmit light only at a wavelength that corresponds to an optical transition to a higher-lying crystal field level of erbium in the preferred sites, whereas it blocks most of the background fluorescence.

Not only the erbium integration, but also the waveguide losses will depend on the implantation and annealing conditions. In general, lower losses are expected for longer annealing and higher temperature. In the following, we thus study the sample implanted at \SI{500}{\celsius}. Based on the manufacturer's loss specification of $\sim \SI{1.5}{\decibel\per\centi\meter}$ (without doping), one would expect a $50\,\%$ reduction of the reflection signal from the longest studied waveguides that have a length of \SI{1}{\centi\meter}. Instead, we observe that the reflection is approximately independent of the waveguide length when taking the considerable fluctuations of the coupling efficiency of the used surface couplers into account. Fig.~\ref{fig:BroadSpectrum}D shows the fluorescence signal as a function of the waveguide length. We find an approximately linear increase. Taken together, this suggests that the loss in the delivered chips is lower than specified, and that the sample fabrication procedure used in this work leads to a lower additional loss as compared to our earlier work that found $\sim \SI{6}{\decibel\per\centi\meter}$ at a comparable overall dose~\cite{gritsch_narrow_2022}. Thus, the added loss obtained in our earlier experiments may not originate from the implantation, but from differences in the fabrication process on implanted rather than undoped samples. However, a definite conclusion and a precise determination of the loss would require analysis of a much larger number of structures than were available for this study.

\section{Magnetic field dependence}

\begin{figure}[htb!]
\includegraphics[width=1.\columnwidth]{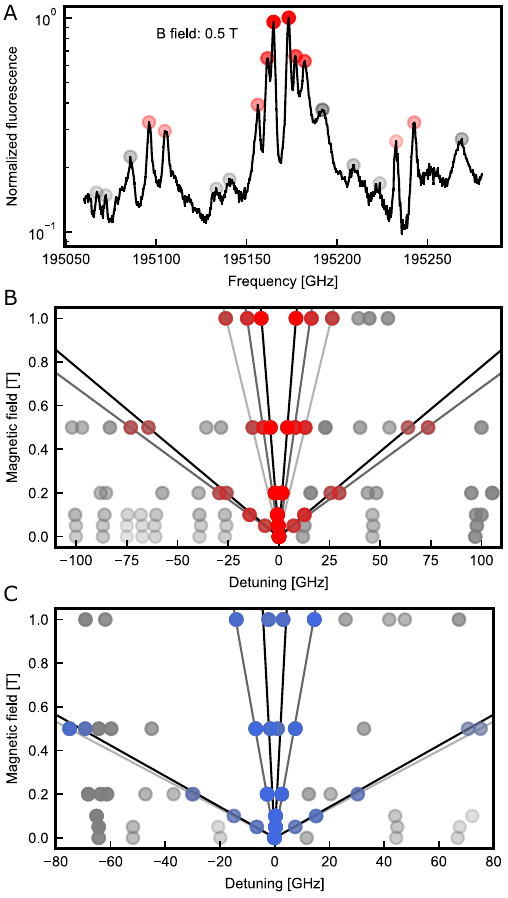}
\caption{     \label{fig:MagneticField}
Magnetic field dependence. A) Splitting of the fluorescence line of erbium in site B when a magnetic field of $\SI{0.5}{\tesla}$ is applied. B, C) Linear splitting (black fit lines) of the observed spin-preserving and spin-flip transitions as a function of the applied magnetic field. Only peaks that split symmetrically around the zero field peak position originate from erbium in the sites A (blue dots) and B (red dots) and are considered in the analysis, while peaks that originate from other sites (gray dots) are ignored. The transparency of the colors indicates the peak amplitude.
}
\end{figure}

After investigating the basic optical properties of the implanted erbium dopants, we now turn to their spin states. While earlier measurements have studied different sites of Er:Si~\cite{vinh_microscopic_2003, yin_optical_2013, yang_zeeman_2022}, this work restricts itself to the recently discovered sites A and B, which have favorable optical properties~\cite{gritsch_narrow_2022}.

In sites of low symmetry, the crystal field lifts the degeneracy of the 4f orbitals of rare-earth dopants. Since erbium is a Kramers ion with an odd number of 4f electrons, its properties in a magnetic field can be modeled as an effective two-level system with an anisotropic g-tensor~\cite{maryasov_spin_2012}, which differs between the ground ($g_g$) and optically excited ($g_e$) states. Thus, the Zeeman effect leads to a splitting of the lines in a magnetic field.

We apply such a field perpendicular to the chip surface, approximately along the [100] crystalline direction. For each dopant, one expects a splitting of the spin-flip transitions, connecting the lower $I_{15/2}$ to the upper $I_{13/2}$ spin level and vice versa, by $\triangle \nu_{sf} \propto |g_g+g_e|  \cdot B$. The spin-preserving lines, in contrast, would be split by $\triangle \nu_{sp} \propto |g_g-g_e| \cdot B$. Fig.~\ref{fig:MagneticField}A shows an example spectrum taken at \SI{0.5}{\tesla}. The peak of erbium in site B (red) is split into six spin-preserving and four resolved weaker spin-flip lines, indicating an asymmetric g-tensor and several magnetic classes that originate from different emitter orientations. Determining the splitting of the lines for several different magnetic field values allows for a clear assignment of the peaks, see panel B) for site B and panel C) for site A. From linear fits we extract $|g_g+g_e|$ and $|g_g-g_e|$ for all observed magnetic classes, as given in Tab. \ref{Table:g_values}. At the applied fields of up to \SI{1}{\tesla}, no deviation from a linear splitting is observed. This changes at higher fields that can induce a mixing of the crystal field levels when their energy would cross \cite{de_boo_high-resolution_2020}.

\begin{table}
\begin{center}
\begin{tabular}{ c || c | c | c } 
Site & $|g_g+g_e|$ & $|g_g-g_e|$  \\ \hline \hline
A & $\SI{21.48 \pm 0.06}{}$ & $\SI{0.6 \pm 0.11}{}$   \\ \hline
A & $\SI{20.26 \pm 0.20}{}$ & $\SI{2.04 \pm 0.05}{}$  \\ \hline
B & $\SI{18.37 \pm 0.12}{}$ & $\SI{1.22 \pm 0.01}{}$  \\ \hline
B & $\SI{20.94 \pm 0.13}{}$ & $\SI{2.23 \pm 0.07}{}$  \\ \hline
B & $-$ & $\SI{3.72 \pm 0.16}{}$ \\
\end{tabular}
\end{center}
\caption{ \label{Table:g_values}
Sum and difference of the effective g-factors of the erbium emitters in sites A and B.
}
\end{table}

\section{Optical coherence}

\begin{figure*}[tb]
\includegraphics[width=2\columnwidth]{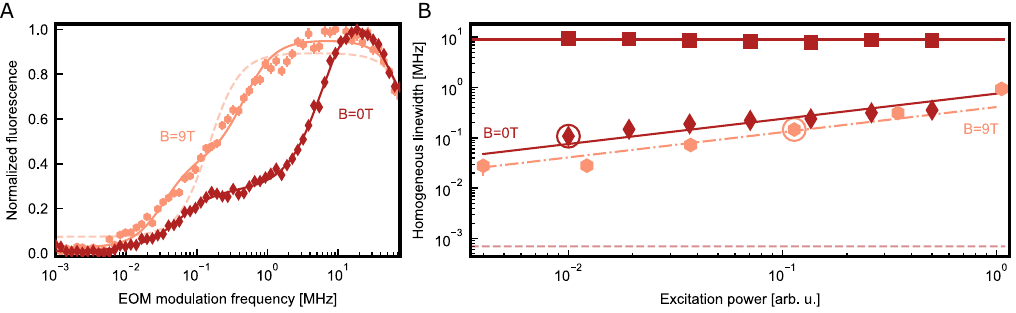}
\caption{\label{fig:HomLine}
A) Measurement of the fluorescence of site B after the laser excitation pulse, modulated to have 27 equidistant lines, as a function of the frequency separation between the lines. The homogeneous linewidth of the dopants can be extracted by Lorentzian fits. In the measurement at zero magnetic field (dark red diamonds), a second decoherence mechanism is present, resulting in a double Lorentzian curve (dark red line). When a magnetic field of \SI{9}{\tesla} is applied (light red hexagons), the width of the second line is reduced (solid fit curve), approaching a single, narrow homogeneous line (dashed fit curve). B) Homogeneous linewidth, extracted from the fit curves, as a function of the excitation laser power. The measurements of panel A are marked by circles. Light red hexagons show the measurement at $\SI{9}{\tesla}$. The measurement at $\SI{0}{\tesla}$ (dark red) shows a narrow (diamonds) and a broader linewidth (squares). In both cases, a reduction of the narrow line is observed down to the lowest powers that give a sufficient signal. But even there, the power broadening is too large to observe a lifetime-limited linewidth (horizontal dashed line).
}
\end{figure*}

After studying the magnetic field dependence of erbium dopants in the preferred sites, we investigate their optical coherence. We use the same technique as in~\cite{weiss_erbium_2021, gritsch_narrow_2022}, where three concatenated electro-optical modulators create a comb of 27 equidistant laser lines in the spectrum of the excitation pulse. Then, the fluorescence is measured as a function of the spectral line separation. Because of saturation caused by transient spectral hole burning, the fluorescence decreases when the modulation is smaller than the homogeneous linewidth. In addition, the signal is reduced when then maximum line separation approaches the inhomogenous linewidth. The linewidths can thus be obtained by fitting the signal decrease with two Lorentzian curves. In home-made samples at zero magnetic field, homogeneous linewidths below $\SI{10}{\kilo\hertz}$ have been observed using this approach.

When performed on commercially fabricated silicon waveguides, the same measurement shows a different result, see Fig.~\ref{fig:HomLine}A (dark red) measured on site B: In addition to a narrow feature with a Lorentzian linewidth of \SI{0.11 \pm 0.01}{\mega\hertz} FWHM, a second increase of the fluorescence with a FWHM of \SI{9 \pm 1}{\mega\hertz} is found. The fact that two timescales are observed indicates that an additional decoherence mechanism is present for some of the dopants in the commercial chips. It can be seen in all investigated samples, independently of the exact implantation and annealing conditions. It also originates from the same site as the narrow feature, which was tested by inserting a narrow-band filter in the detection path that only transmits light on a transition to a higher-lying crystal field level.

We attribute the additional decoherence that is not observed in home-made chips to paramagnetic impurities, which may already be present in the starting wafer or be added during the commercial fabrication process. To further investigate this, we apply a large magnetic field of up to \SI{9}{\tesla}. At the chip temperature of $<\SI{2}{\kelvin}$, the spin of both the erbium dopants and all paramagnetic impurities will then be frozen to the ground state, such that magnetic field fluctuations and the resulting decoherence can be strongly reduced~\cite{bottger_effects_2009}. In our samples, the second line is slightly reduced at $\SI{1}{\tesla}$, and almost fully disappears at $\gtrsim \SI{4}{\tesla}$, where the paramagnetic spins are frozen, as shown in Fig.~\ref{fig:HomLine}A for a field of $\SI{9}{\tesla}$ (light red). This supports our hypothesis that paramagnetic impurities are the origin of the added decoherence.

We further investigate the scaling of the homogeneous linewidths of erbium in both sites with the applied laser power in Fig. \ref{fig:HomLine}B. As expected, we observe a drop of the homogeneous linewidth down to the lowest power that gives a sufficient signal, both without magnetic field (dark red) and at $\SI{9}{\tesla}$ (light red). With the field applied, we extract a value of \SI{0.028 \pm 0.01}{\mega\hertz} as an upper bound to the homogeneous linewidth. Thus, we can almost recover the narrow homogeneous linewidths observed in home-made samples \cite{gritsch_narrow_2022} by applying a magnetic field of sufficient strength. The obtained linewidth, still limited by power broadening, is within an order of magnitude from the lifetime limit that stems from the exponential decay of the emitters, with a measured timescale of $\SI{197 \pm 2}{\micro\second}$ and $\SI{227 \pm 3}{\micro\second}$ for site A and B, respectively, in the used waveguide geometry. With no field applied, we observe a similar behavior of the narrow feature (diamonds), but a constant linewidth of \SI{9 \pm 1}{\mega\hertz} for the additional line (squares). This is expected, as the dipolar interaction with paramagnetic impurities does not depend on the applied laser power.

\section{Summary and outlook}

In summary, we have shown that photon emitters at a telecommunication wavelength can be reliably integrated into silicon nanostructures that are commercially fabricated on a wafer-scale. In spite of added impurities, the excellent optical coherence of erbium dopants~\cite{gritsch_narrow_2022} can be re-established in these devices by applying a large magnetic field. Thus, emission of coherent single photons is expected in nanophotonic resonators that can reduce the lifetime sufficiently \cite{gritsch_purcell_2023}. In comparison to other emitters in silicon that feature shorter lifetimes~\cite{redjem_single_2020, durand_broad_2021, higginbottom_optical_2022, prabhu_individually_2023, redjem_all-silicon_2023}, erbium does not suffer from non-radiative decay, and exhibits much narrower homogeneous, inhomogenous and spectral diffusion linewidths \cite{gritsch_narrow_2022, gritsch_purcell_2023}. This is a prerequisite for using the techniques of spectral holeburning and photon echoes to realize efficient quantum memories~\cite{tittel_photon-echo_2010}. To this end, the achieved coherence should enable storage on a microsecond timescale using the electronic spin transitions investigated in this work, similar to earlier work with YSO \cite{lauritzen_telecommunication-wavelength_2010, craiciu_multifunctional_2021, liu_-demand_2022}. Much longer coherence times, likely exceeding seconds, can then be achieved by storage in the hyperfine states of the $^{167}\text{Er}$ isotope under high magnetic fields \cite{rancic_coherence_2018}. The limitation of the storage time by the coupling to the bath of nuclear spins can be avoided in isotopically purified silicon \cite{mazzocchi_99992_2019} or silicon-on-insulator \cite{liu_28silicon--insulator_2022} chips. This may pave the way to hour-long coherence, as demonstrated for donors without an optical interface \cite{saeedi_room-temperature_2013}. To implement such memories, improving the purity of the starting wafer and optimizing the fabrication process based on the fluorescence signal may be advantageous to increase the optical depth, and to reduce both the background and the decoherence at zero field. With this, implementing an efficient quantum memory in the demonstrated industrially fabricated samples may aid the implementation of scalable optical quantum computers by photon synchronization~\cite{nunn_enhancing_2013, makino_synchronization_2016}, and allow for the connection of superconducting quantum computers via microwave-to-telecom conversion \cite{obrien_interfacing_2014}. In addition, our system may allow for the implementation of global quantum networks via dedicated quantum repeater protocols~\cite{sinclair_spectral_2014}.

\section{Acknowledgments}
This project received funding from the European Research Council (ERC) under the European Union's Horizon 2020 research and innovation program (grant agreement No 757772), from the Deutsche Forschungsgemeinschaft (DFG, German Research Foundation) under the German Universities Excellence Initiative - EXC-2111 - 390814868 and via the project RE 3967/1, and from the German Federal Ministry of Education and Research (BMBF) via the grant agreements No 13N15907 and 16KISQ046.

\bibliographystyle{naturemagsr.bst}
\bibliography{bibliography.bib}

\end{document}